\documentclass[preprint,12pt]{elsarticle}

\usepackage{amssymb}
\usepackage{amsmath}
\usepackage{amsthm}

\usepackage{algorithm}
\usepackage{algorithmicx}
\usepackage{algpseudocode}

\newtheorem{theorem}{Theorem}

\newtheorem{corollary}{Corollary}
\newtheorem{proposition}{Proposition}

\theoremstyle{definition}
\newtheorem{definition}{Definition}

\begin{document}

\begin{frontmatter}



\title{Polynomial-Solvability of $\cal{NP}$-class Problems }


\author{Anatoly Panyukov}

\address{paniukovav@susu.ru}

\begin{abstract}
Let Hamiltonian complement of the graph $G = (V(G),\, E(G))$ be the minimal cardinality set $H(G)\subset V(G)\times V(G)$ such that graph $(V(G),\, E(G)\cup H(G))$ is a Hamiltonian one. Possibility to recognize the cardinality of Hamiltonian complement $H(G)$ based on reduction to solving the linear programming problem is presented in this paper.
Polynomial solvability of ${\cal NP}$-class follows from the fact that \mbox{$ {\cal NP}$-complete problem} "Hamiltonian circute" is special case of the problem under consideration.
\end{abstract}


\end{frontmatter}


\section{Introduction}
One of the top problems in the theory of algorithms is \mbox{${\cal P} \ vs \ {\cal NP}$}~\cite{ClayMath}, here ${\cal P}$ is the class of problems solvable by algorithms with polynomial computational complexity by deterministic machines,
${\cal NP}$ is the class of recognition problems solvable by nondeterministic machines in polynomial time.

The foundation of ${\cal NP}$-completness theory was laid down by S.~Cook~\cite{Cook1971}. He~introduced the class ${\cal NP}$ of recognition problems, and the concepts of polynomial reducibility and ${\cal NP}$-complete problem \cite{CookSA1983}.
\mbox{${\cal NP}$-completeness} of a wide range of recognition problems was proved. In particular, this also applies to the "Hamiltonian circuit" problem \cite{Karp1972}.   The most complete guide to the theory of \mbox{${\cal NP}$-completeness} is the monograph \cite{Garey}.

Currently, the question "$Are\; {\cal NP}$-complete problems difficult to solve?"  is considered one of the main issues of modern mathematics \cite{ClayMath}. It is known that the proof of the possibility to solve at least one ${\cal NP}$-complete problem by a polynomial algorithm be a proof of the coincidence of the classes ${\cal P}$ and ${\cal NP}$.

The polynomial method for reducing the problem of recognizing the cardinality of a Hamiltonian complementation of a graph to solve the linear programming problem is proposed in this preprint. It proves polynomial-solvability of all problems of \mbox{${\cal NP}$-class}.

\section*{Hamiltonian covering of the graph}
\label{Int}
\begin{definition}\label{def:HCow}
Let Hamiltonian complementation of graph $G=()V(G),\,E(G)$ be minimal cardinality set $H(G)\subset V(G)\times V(G)$ such that the graph $(V(G),\, E(G)\cup H(G))$ is a Hamiltonian one. 
\end{definition}
Obviously, the problem to recognize the Hamiltonian complement cardinality is a generalization of the ${\cal NP}$-complete problem "Hamiltonian circuit"\; \cite{Karp1972}.
Consequently, the existence of a polynomial algorithm for the problem of recognizing the Hamiltonian complement cardinality is a proof of the polynomial solvability of ${\cal NP}$-class problems.

\section*{Chain location problem for graph $G$}
\label{sec-look}

Let $C$ be a chain with the set of vertices
$ V(C)=\{c_1,c_2,\dots,c_{n}\}$
and the set of edges
$
 E(C)=\left\{ \{c_i,c_{i+1}\}:\
 i=1,2,\dots ,n \right\} .
$

The challenge to recognize the existence of a Hamiltonian path in graph $ G $ is equivalent to the problem of recognition of  the bijection
\[
 \varphi:\ V(C)\leftrightarrow V(G):\ \{\varphi (c_i),\varphi (c_{(i+1)})  \}\in E(G),\quad i=1,2,\dots ,n-1.
\]
existence.

Task of recognizing existence of the bijection can be represented as a Boolean quadratic programming problem.
Indeed, let us define  $$x=\{x^i_v:\ i=1,2,\dots,n;\ v\in V(G)\}$$ as following
\begin{equation}
\label{Equ:xv}  x^i_v=\delta_v^{\varphi (c_i)}=\left\{ \begin{array}{lc}
1, & \text{ if } \varphi(c_i)=v, \\
0, & \text{ if } \varphi(c_i)\not = v,
\end{array} \right. \qquad  i=1,2,\dots , n,\ v\in V(G).
\end{equation}

It is clear that the definition of elements of the set $x$ by (\ref{Equ:xv}) establishes a one-to-one correspondence between the mapping $\varphi :\ V(C)\rightarrow V(G)$ and a point of the unit cube $\{0,1\}^{n^2}$. {\em Unambiguous}  mappings $\varphi $ correspond to the vertices of the unit cube, belonging to the set
\begin{equation}\label{Equ:OneVal}
	D_1=\left\{x:\ \sum_{v\in V(G)} x^i_v = 1,\, i=1,2,\dots , n,\, x\ge 0 \right\}
\end{equation}
because restrictions (\ref{Equ:OneVal}) of the problem are the requirement that every element of $c_i\in V(C)$ receives exactly one destination.

{\em Surjective } mappings of $\varphi $ correspond to the vertices of the unit cube belonging to the set
\begin{equation}\label{Equ:Inj}
	D_2=\left\{x:\ \sum_{i=1}^{n} x^i_v = 1,\, v\in V(G) \right\}
\end{equation}
because group of restrictions (\ref{Equ:Inj}) is the requirement that each element $v\in V(G)$ is assigned a unique element $c_i\in V(C)$. {\em Bijective} mappings of $\varphi $ correspond to the vertices of the unit cube $$D_3=\left\{ x^i_v\in \{0,\,1\}:\ i=1, 2, \dots , n-1, \; v\in V(G) \right\}$$ belonging to the set $D_1$, and the set $D_2$, i.e. all elements of the set $$D=D_1\cap D_2\cap D_3.$$

Let us consider the Boolean optimization problem
\begin{equation}\label{Equ:Loss}
	F(x)=\sum_{i=1}^{n-1}\left(\sum_{v,u\in V(G):\, \{v,u\}\not\in {E}({G})} x^i_v x^{i+1}_u\right)  \ \to \ \min_{x\in D}.
\end{equation}
The value of the objective function $F(x)$ is equal to the number of edges in the set $\overline{E({G})}=[V(G)]^2\setminus E(G)$  that is image $\varphi(C)$ of the arranged chain $C$ .
\begin{proposition}\label{prop:1}
Let $x^*$ be an optimal solution of the problem~(\ref{Equ:Loss}). The graph $G$ contains a Hamiltonian path if and only if $F(x^*)=0$.
\end{proposition}
\begin{proposition}\label{prop:2}
Let $x^*$ be an optimal solution of the problem~(\ref{Equ:Loss}), then $|H(G)|=F(x^*)$.
\end{proposition}

\section*{Presentation of the problem (\ref{Equ:Loss}) as the ILP problem}
Let us introduce boolean variables
\begin{equation} \label{yxx}
y^{(i,\, i+1)}_{(u,\, v)}=x^i_ux^{i+1}_v,\; i=1, 2, \dots , n-1, \; u, v \in V(G).
\end{equation}
It follows from (\ref{Equ:OneVal}) and (\ref{yxx}) that
\[
\sum_{v\in V(G)}y^{(i,\, i+1)}_{(u,\, v)} = x^i_u \sum_{v\in V(G)}x^{i+1}_v=x^i_u,
\]
\[
\sum_{u\in V(G)}y^{(i,\, i+1)}_{(u,\, v)} = x^{i+1}_v \sum_{u\in V(G)}x^{i}_u=x^{i+1}_v.
\]
Let us consider the boolean linear programming problem
\begin{equation} \label{F_q}
F_Q(x,y)=\sum_{i=1}^{n-1} \sum_{v,u\in V(G)}y^{(i,\, i+1)}_{(v,\,u)}\chi _{\overline{E(G)}}(\{u,v\})\ \rightarrow
\min_{\substack{ x\in D, \\ (x,y)\in M}}
\end{equation}
where
\begin{multline} \label{Equ:M_Q}
M=\tilde{M} \bigcap \\ \left[ D_3 \times \left\{ y^{(i, \, i+1)}_{(u, \, v)}\in\{0,1\}:\ i=1,2, \dots n-1, \; u, v \in V(G) \right\} \right],
\end{multline}
\begin{multline} \label{Equ:M_Q_R}
\tilde{M}=\left\{(x,y) \geq 0\:\ \displaystyle \sum_{w\in V(G)} y^{(i,\, i+1)}_{(u,\,w)} =x^i_u,\right. \\ \left.  \displaystyle \sum_{w\in V(G)} y^{(i,\, i+1)}_{(w,\,v)}=x^{i+1}_v,\; i=1,2, \dots n-1, \;  u,v\in V(G) \right\}.
\end{multline}
Note that the system $\tilde{M}$ of restrictions differs from the system $M$ of restrictions by the lack of integrality conditions. Later the set $\tilde{M}$ is used for construction of relaxed problems.

It follows from (\ref{F_q}) that
\begin{proposition}\label{prop:IntLP}
A necessary and sufficient condition for optimality of the problem (\ref{Equ:Loss}) solution $x^*$  is the optimality of the  problem~(\ref{F_q}) solution $(x^*,\,y^*)$.
\end{proposition}
\begin{proof}
Let $x^{(1)}$ be the optimal solution of problem~(\ref{Equ:Loss}). Let us define $y^{(1)}$ in accordance with (\ref{yxx}). Then $(x^{(1)},\,y^{(1)})$ is a valid problem (\ref{F_q}) solution. Conversely, if $(x^{(2)},\,y^{(2)})$ is the problem~(\ref{F_q}) optimal solution, then (\ref{F_q}) -- (\ref{Equ:M_Q_R}) imply that $x^{(2)}$ is a valid problem (\ref{Equ:Loss}) solution. In this way,
\begin{equation}
\label{EQU}
F(x^{(1)})=F_Q((x^{(1)},\,y^{(1)}))\geq F_Q((x^{(2)},\,y^{(2)}))=F(x^{(2)})\geq F(x^{(1)}).
\end{equation}
Consequently,  equalities hold in the chain (\ref{EQU}), i.e. all considered in the proof solutions are optimal solutions of the corresponding problems.
\end{proof}

\section*{Relaxed chain location problem}
Let us consider the relaxation of  problem (\ref{F_q})
\begin{equation} \label{F_Qr}
F_Q(x,y)= \sum_{i=1}^{n-1} \sum_{v,u\in V(G)}y^{(i,\, i+1)}_{(v,\,u)}\chi _{\overline{E(G)}}(\{u,v\})\ \rightarrow
\min_{\substack{ x\in D_1\cap D_2\, \\ (x,y)\in \tilde{M}}}.
\end{equation}
Constraints of relaxed  problem (\ref{F_Qr}) are different from the constraints of source problem (\ref{F_q} by absence of integrity  restriction $x\in D_3$.

The dual to (\ref{F_Qr})) problem is the following
\begin{equation}\label{dQAP1}
F^*_Q(\xi,\eta,\lambda )=\sum _{i=1}^n \xi_i + \sum _{v\in V(G)}\lambda _v\rightarrow \max_{(\xi,\eta,\lambda) \in \tilde{M}_Q^*}
\end{equation}
where the feasible set
\begin{multline}\label{equ:xi}
\tilde{M}^*
=\left\{(\xi, \lambda, \eta):
\xi_1 - \eta^{(1,\, 2)}_v\leq \lambda_v,\; \xi_n - \eta^{(n,\,n-1)}_v\leq \lambda_v, \; v\in V(G),
\right. \\ \left.
\xi_{i} - \eta^{(i,\, i+1)}_v - \eta^{(i,\, i-1)}_v		\leq \lambda_v, \qquad 2\leq i\leq n-1 ,\;  v\in V(G),
\right. \\ \left.
\eta^{(i,\,i+1)}_v + \eta^{(i+1,\, i)}_u \leq \chi_{\overline{E(G)}}(\{u, v\}), \;
1\leq i\leq n-1 ,\;   u,v\in V(G).
 \right\},
\end{multline}
here variables $\xi $ correspond to the restrictions of the set $D_1$, variables $\lambda $ correspond to restrictions of the set $ D_2 $, variables $\eta $ correspond to restrictions of the set $\tilde{M}$.

Let us introduce the subset
\[
\tilde{L}^* = \{ (\xi, \lambda, \eta)\in \tilde{M}^* :\  \sum_{v\in V}\lambda_v  = 0 \} \subset \tilde{M}^*
\]
and the problem
\begin{equation}\label{drveblp2_}
\tilde F_{Q_L}^*(\xi,\lambda, \eta)=\sum _{i=1}^n\xi^i  \rightarrow \max_{(\xi, \lambda, \eta)\in \tilde{L}^*}.
\end{equation}
\begin{proposition}\label{prop:DuaLP}
All optimal solutions of problem (\ref{drveblp2_}) are optimal solutions of problem (\ref{dQAP1}).
\end{proposition}
\begin{proof}
Let us put
\[
(\xi^{\pi}=\xi+\hat{\lambda},\; \lambda^{\pi}=\lambda-\hat{\lambda},\; \eta^{\pi}=\eta),\quad \hat{\lambda}=\left(\frac{\sum_{v\in V}{\lambda_v}}{|V|}\right){\sf e}.
\]
Obviously,
\[ (\xi^{\pi},\, \lambda^{\pi},\, \eta^{\pi}) \in \tilde{M}^*,\qquad \sum _{v\in V}\lambda^{\pi}_v = 0.\]
So $(\xi^{\pi},\, \lambda^{\pi},\, \eta^{\pi}) \in \tilde{L}^*$,
and
\[
\sum _{i=1}^n{(\xi^{\pi})}^i = \sum _{v\in V}\lambda_v + \sum _{i=1}^n\xi^i .
\]
Proposition \ref{prop:DuaLP} is proved.
\end{proof}
\begin{theorem}\label{def:HCow}
The set of optimal solutions of the relaxed problem (\ref{F_Qr}) contains an integer solution.
\end{theorem}
\begin{proof}
Let
\[
(\xi^*, \lambda^*, \eta^*)= \arg  \max_{(\xi, \lambda, \eta)\in
\tilde{M}^*}\left( \sum _{v\in V}\lambda_v + \sum _{i=1}^n\xi^i \right)
\]
be optimal solutions of problem (\ref{dQAP1}).

It is easy to see that for fixed values of dual variables $ \lambda^* $ the problem (\ref{dQAP1}) turns out to be a dual problem for the problem $\Theta _W (\lambda^*)$:
\begin{multline}\label{Equ:WeberPrb_}
	F_W(\lambda^*)(x,y)=-\sum_{i\in V(C),\, v\in V(G)}\lambda^*_v  x^i_v + \\
	 \sum_{ (i,i^+)\in E(C)} \sum_{v,u\in V(G)}y^{(i,\,i^+)}_{(v,\,u)}\chi _{\overline{E(G)}}(\{u,v\})  \ \to \ \min_{\substack{ x\in D_1 \\ (x,y)\in \tilde{M}}}.
\end{multline}
Here, in contrast to problem (\ref{Equ:Loss}), the surjective condition (\ref{Equ:Inj}) is absent, and the cost $\lambda^*_v$ of placing of vertices $i\in V(C)$ onto vertices $v\in V(G)$ is added.
\begin{proposition}\label{prop:WPrb}
All optimal solutions of problem (\ref{Equ:WeberPrb_}) belong to the convex hull of its integer optimal solutions.
\end{proposition}
\begin{proof}
Let as introduce graph $\cal{G}$ with vertexes set
\begin{equation*}
V({\cal{G}})=\left\{ v_0,\, v_{n+1} \right\} \bigcup \left[ V(G) \times \left\{1,2,\dots , n \right\} \right]
\end{equation*}
and edges set
\begin{multline*}
E({\cal{G}})=\left[\bigcup_{v\in V(G)}\left\{ \left(v_0,\,(v,1) \right) \right\} \right] \bigcup \left[\bigcup_{v\in V(G)}\left\{ \left((v,n),\,v_{n+1} \right) \right\} \right] \\ \bigcup \left[\bigcup_{i=1}^{n-1}\left(\bigcup_{u,v\in V(G)}\left\{  \left((u,i),\,(v,i+1) \right) \right\} \right)\right] .
\end{multline*}
Problem (\ref{Equ:WeberPrb_}) in terms of ${\cal G}$ is the problem of finding the minimum weight path between vertices $ v_0 $ and $ v_{n + 1} $, provided that the weights of the vertices $ (v, i) \in V ({\cal{G}}) $ equal to $(-\lambda_v ) $, the weights of the edges $\left\{ (u, i),\,(v, i + 1) \right\} \in E({\cal{G}}) $ are equal to $\chi_{\overline{E (G)}}(\{u, v\}) $, the weights of the vertices $ v_0, v_{n + 1} $ and the edges incident to them are zero.

We can find the set
\begin{equation} \label{Equ:WSltn}
S=\left\{
(x^o,y^o)^{(k)}=\arg \min_{(x,y)\in \tilde{M}}F_W(\lambda^*)(x,y),\; k=1,\,2,\,\dots ,K
\right\}
\end{equation}
of all optimal solutions using the known shortest path algorithms. Obviously, these solutions satisfy the condition $x^o\in D_1\cap D_3$.

The restriction matrix of the problem $F_W(\lambda^*)$ is completely unimodular. Consequently the set of all optimal solutions of the problem $F_W(\lambda^*)$ represents the convex hull of $\mbox{\sf Conv}\,S $ of the set of its optimal integer solutions (that is defining by all optimal paths between the vertices $v_0$ and $v_{ n + 1}$).
\end{proof}

Let us show that the chain of relations
\begin{multline}\label{Equ:Chain}
\min_{\substack{ x\in D_1\cap D_2, \\ (x,y)\in \tilde{M}}}F_Q(x,y)=
\max_{(\xi, \lambda, \eta)\in \tilde{M}^*}\tilde F_Q^*(\xi,\lambda, \eta) = \\
= \tilde F_Q^*(\xi^*,\lambda^*, \eta^*) =
\max_{(\xi, \eta):\, (\xi,\lambda^*, \eta)\in \tilde{L}^*}\tilde F_Q^*(\xi,\lambda^*, \eta) = \\ = \min_{\substack{ x\in D_1, \\ (x,y)\in \tilde{M}}}F_W(\lambda^*)(x,y) =
\min_{\substack{ x\in D_1\cap D_3, \\ (x,y)\in \tilde{M}}}F_W(\lambda^*)(x,y) \leq \\ \leq
\min_{\substack{ x\in D_1\cap D_2, \\ (x,y)\in \tilde{M}}}F_W(\lambda^*)(x,y)
= \min_{\substack{ x\in D_1\cap D_2, \\ (x,y)\in \tilde{M}}}F_Q(x,y).
\end{multline}
is hold.

The first equality is a consequence of the linear programming first duality theorem for the pair of mutually dual problems (\ref{F_Qr}) and (\ref{dQAP1}).
The next two equalities follow from the theorem conditions and Proposition  \ref{prop:DuaLP}. The fourth equality is a consequence of the first linear programming duality theorem for a pair of mutually dual problems (\ref{Equ:WeberPrb_}) and (\ref{dQAP1}). The fifth equality is a consequence of the Proposition \ref{prop:WPrb}.
Inequality is the result of the restriction of an admissible set for a minimum function.
The last equality follows from the definition of the set $\tilde{L}^*$ and the inclusion $x\in D_2$.

So, the optimal value of problem (\ref{Equ:WeberPrb_}) for $\lambda^*\in \tilde{L}^*$  coincides with the optimal value of problems (\ref{F_Qr}) and (\ref{dQAP1}).
In accordance with the Proposition~\ref{prop:WPrb}, the optimal solution $(x^*,y^*)$ belongs to the set $D_2\cap \mbox{\sf Conv}\,S $, but this is possible only if  $S\subset D_1\cap D_2\cap D_3$.
Indeed, the inclusion of $S\subset D_1\cap D_3$ is a consequence of Proposition~\ref{prop:WPrb}. The assumption $S \nsubseteq D_2\cap D_3$ contradicts to optimality of $\lambda^*$.
Theorem now follows.
\end{proof}
The proof of the theorem establishes the existence of an optimal integer solution of the problem (\ref{F_Qr}), but does not give an algorithm for finding this solution. Nevertheless, the existence of an optimal solution of the problem (\ref{F_Qr}) makes it possible to determine the Hamiltonian complementation cardinality.

\begin{corollary}
The optimal value of the problem (\ref{dQAP1}) for the graph $G$ is equal to the cardinality of the Hamiltonian complementation $|H(G)|$.
\end{corollary}

\section*{Conclusion}
Problem (\ref{dQAP1}) represents a linear programming problem to solve which the polynomial algorithms \cite{Khachian} are known. The recognizing problem of the presence of a Hamiltonian circuit in a graph belongs to the class ${\cal NP}$ \cite{Karp1972}. Hence, we have proved the theorem
\begin{theorem}
All problems of ${\cal NP}$ class are polynomial-solvable with deterministic machine.
\end{theorem}

\bibliographystyle{elsarticle-num}
\bibliography{PAV_DB_Eng}



\end{document}